\documentclass[submission, Proceedings]{SciPost}

\hypersetup{
    colorlinks,
    linkcolor={red!50!black},
    citecolor={blue!50!black},
    urlcolor={blue!80!black}
}

\usepackage[bitstream-charter]{mathdesign}
\usepackage{subfig}
\usepackage{comment}
\DeclareSymbolFont{usualmathcal}{OMS}{cmsy}{m}{n}
\DeclareSymbolFontAlphabet{\mathcal}{usualmathcal}

\begin{document}

%\nolinenumbers

\begin{center}{\Large \textbf{
ATLAS results on charmonium production\\
}}\end{center}

% TODO: write the author list here. Use initials + surname format.
% Separate subsequent authors by a comma, omit comma at the end of the list.
% Mark the corresponding author with a superscript *.
\begin{center}
V. Kartvelishvili%\textsuperscript%{1$\star$}
\\
on behalf of the ATLAS collaboration
\end{center}

\begin{center}
%{\bf 1}
%on behalf of the ATLAS collaboration
%\\
%\ \ \  
%\\ 
(Lancaster University, Lancaster, UK)
\\
{\textit {Email: v.kartvelishvili@cern.ch}}
\end{center}

\begin{center}
\today
\end{center}

% For convenience during refereeing (optional),
% you can turn on line numbers by uncommenting the next line:
%\linenumbers

\definecolor{palegray}{gray}{0.95}
\begin{center}
\colorbox{palegray}{
  \begin{tabular}{rr}
  \begin{minipage}{0.1\textwidth}
    \includegraphics[width=22mm]{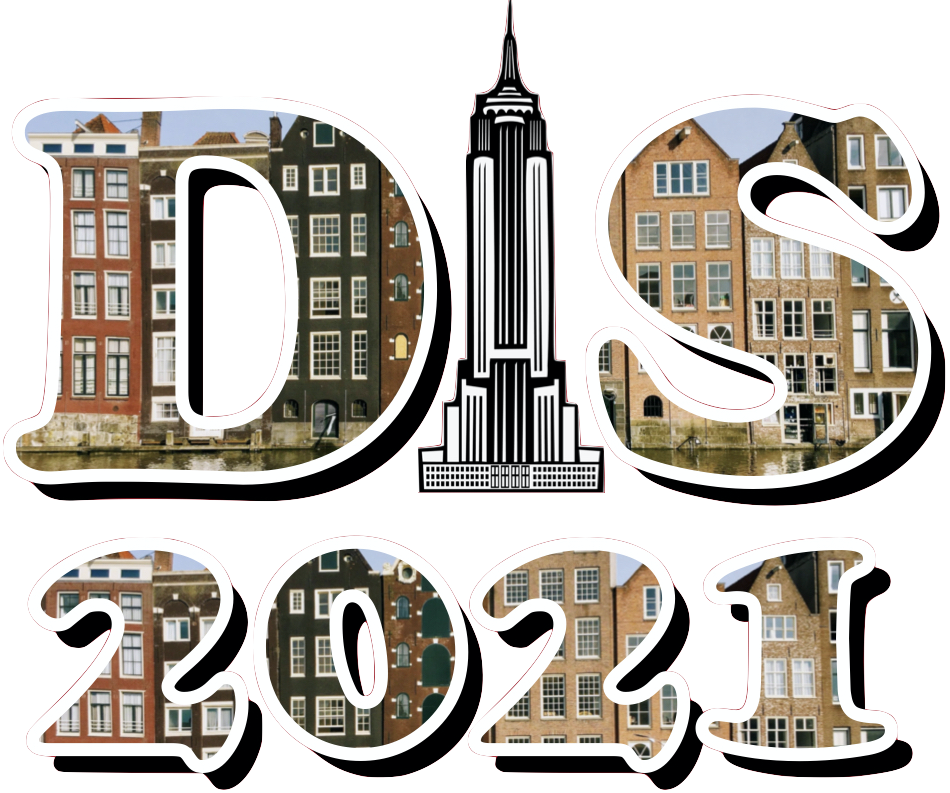}
  \end{minipage}
  &
  \begin{minipage}{0.75\textwidth}
    \begin{center}
    {\it Proceedings for the XXVIII International Workshop\\ on Deep-Inelastic Scattering and
Related Subjects,}\\
    {\it Stony Brook University, New York, USA, 12-16 April 2021} \\
    \doi{10.21468/SciPostPhysProc.?}\\
    \end{center}
  \end{minipage}
\end{tabular}
}
\end{center}

\section*{Abstract}
{\bf
% TODO: write your abstract here.
An overview is given of the results on charmonium production obtained by the ATLAS collaboration at the LHC. Recent preliminary measurements covering prompt and non-prompt production of $J/\psi$ and $\psi(2S)$ states in the range of high transverse momenta from 60 to 360~GeV are highlighted. 
}

\section{Introduction}
\label{sec:intro}
% TODO: write your article here.

Charmonium is an important object of study in QCD at the boundary of small and large momentum scales. Apart from a major role in understanding its production mechanisms in hadronic -- and other -- collisions, various charmonium and bottomonium states are increasingly used as tools for investigating other phenomena, such as studies of quark-gluon plasma in heavy ion collisions, separation of single- and double-parton scattering processes, measurement of transverse-momentum-dependent parton distributions, and many others.

ATLAS~\cite{ATLAS:2008xda} has a long history of charmonium production measurements, covering both inclusive (see, e.g., 
\cite{ATLAS:2015zdw,ATLAS:2014ala,TheATLAScollaboration:2015anj,ATLAS:2016kwu}) and associative
\cite{ATLAS:2014yjd,ATLAS:2014ofp,ATLAS:2016ydt,ATLAS:2019jzd} production processes.

\section{Charmonium at high $p_T$}

The main topic of this report is the measurement of the double-differential cross section of $J/\psi$ and $\psi(2S)$ production at ATLAS\cite{ATLAS:2019ilf}. The measurement covers the range of absolute rapidity $|y|$ from 0 to 2 in 3 bins, and the range of high transverse momenta, $p_T$, from 60 to 360~GeV in 12 bins, which goes well beyond the highest charmonium $p_T$ ever reached before. Data are collected by the ATLAS detector in proton-proton collisions at the centre-of-mass energy of 13~TeV, using a single-muon trigger with a 50~GeV threshold, over the full Run II period with an integrated luminosity of 139~fb$^{-1}$. 

\begin{figure}[h]
%\begin{figure}[htbp!]
\centering
\subfloat[]{\includegraphics[height = 5.4cm, keepaspectratio]{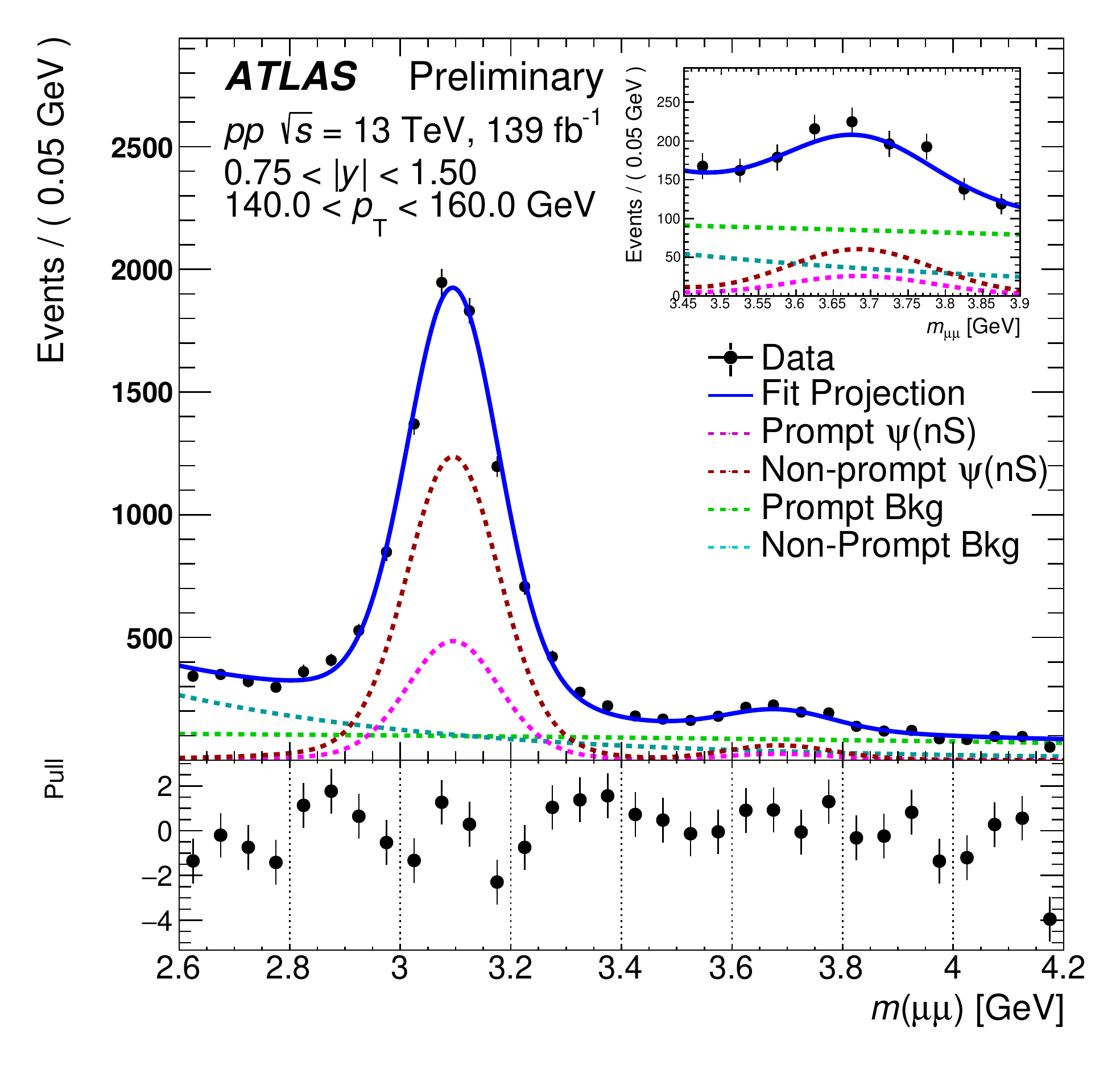}}
\subfloat[]{\includegraphics[height = 5.4cm, keepaspectratio]{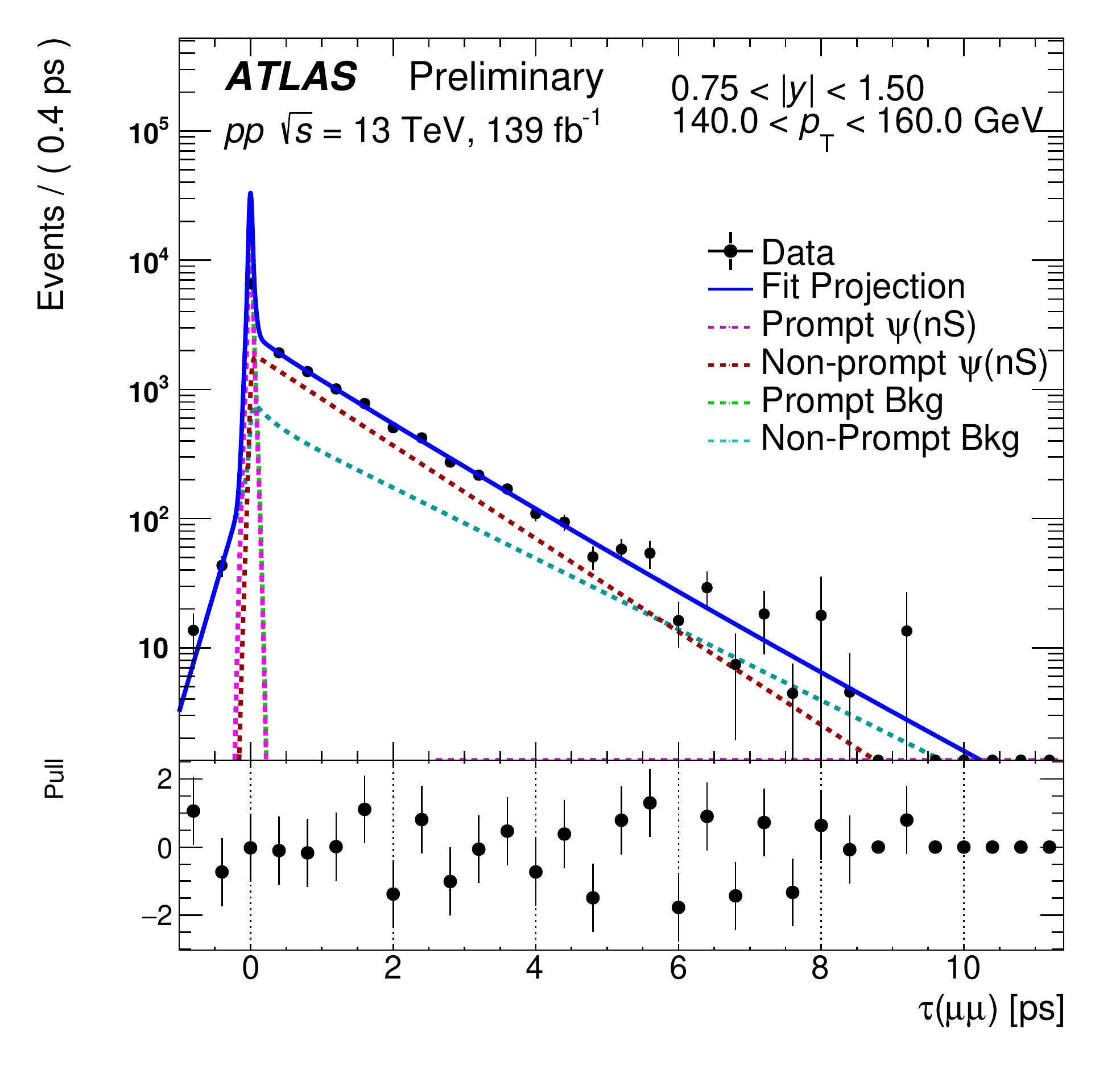}}
\caption{
Dimuon invariant mass (a) and pseudo-proper lifetime (b) projections of the fit result for a bin with $0.75 < |y| < 1.5$ and $140$~GeV$ < p_T < 160$~GeV~\cite{ATLAS:2019ilf}.
}
\label{fig:2dfits}
\end{figure}

An important feature of charmonium production measurements at LHC is the ability of experimental separation of prompt (QCD sources) and non-prompt ($B$ hadron decays) mechanisms. This exploits the property of $B \to J/\psi X$ decays, where the $J/\psi$  tends to carry a fixed fraction of the parent’s momentum. In ATLAS the distribution of the pseudo-proper lifetime $\tau={m\, L_{xy}}/{c\, p_T}$ is used for this purpose, where $m$ is the dimuon mass, $L_{xy}$ the decay length of the dimuon system and $c$ is the speed of light. Promptly produced charmonia have $\tau$ consistent with zero within experimental resolution, while non-prompt events form a (quazi-) exponential tail.
In each $p_T$-$y$ bin, a 2D maximum likelihood fit was performed in the dimuon invariant mass and pseudo-proper lifetime, to determine the yields of prompt and non-prompt $J/\psi$ and $\psi(2S)$ mesons (see 
Fig.~\ref{fig:2dfits} for fit projections in one of the bins). 
All relevant corrections for acceptance, trigger and reconstruction efficiencies have been applied and double-differential cross sections, non-prompt fractions and  $\psi(2S)$-to- $J/\psi$ production ratios were measured.

\begin{figure}[h]
%\begin{figure}[htbp!]
\centering
\subfloat[]{\includegraphics[height = 5.0cm, keepaspectratio]{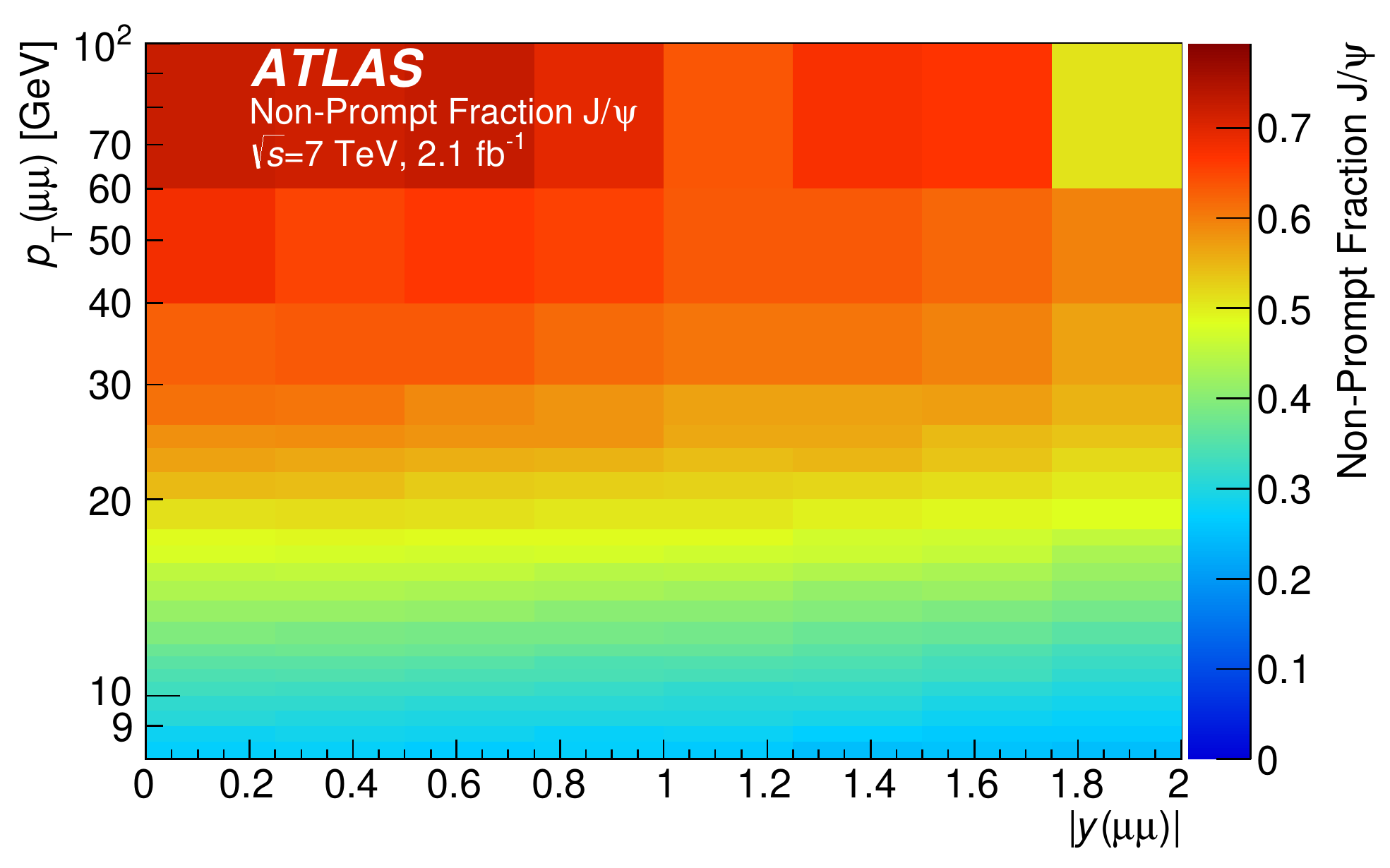}}
\subfloat[]{\includegraphics[height = 5.0cm, keepaspectratio]{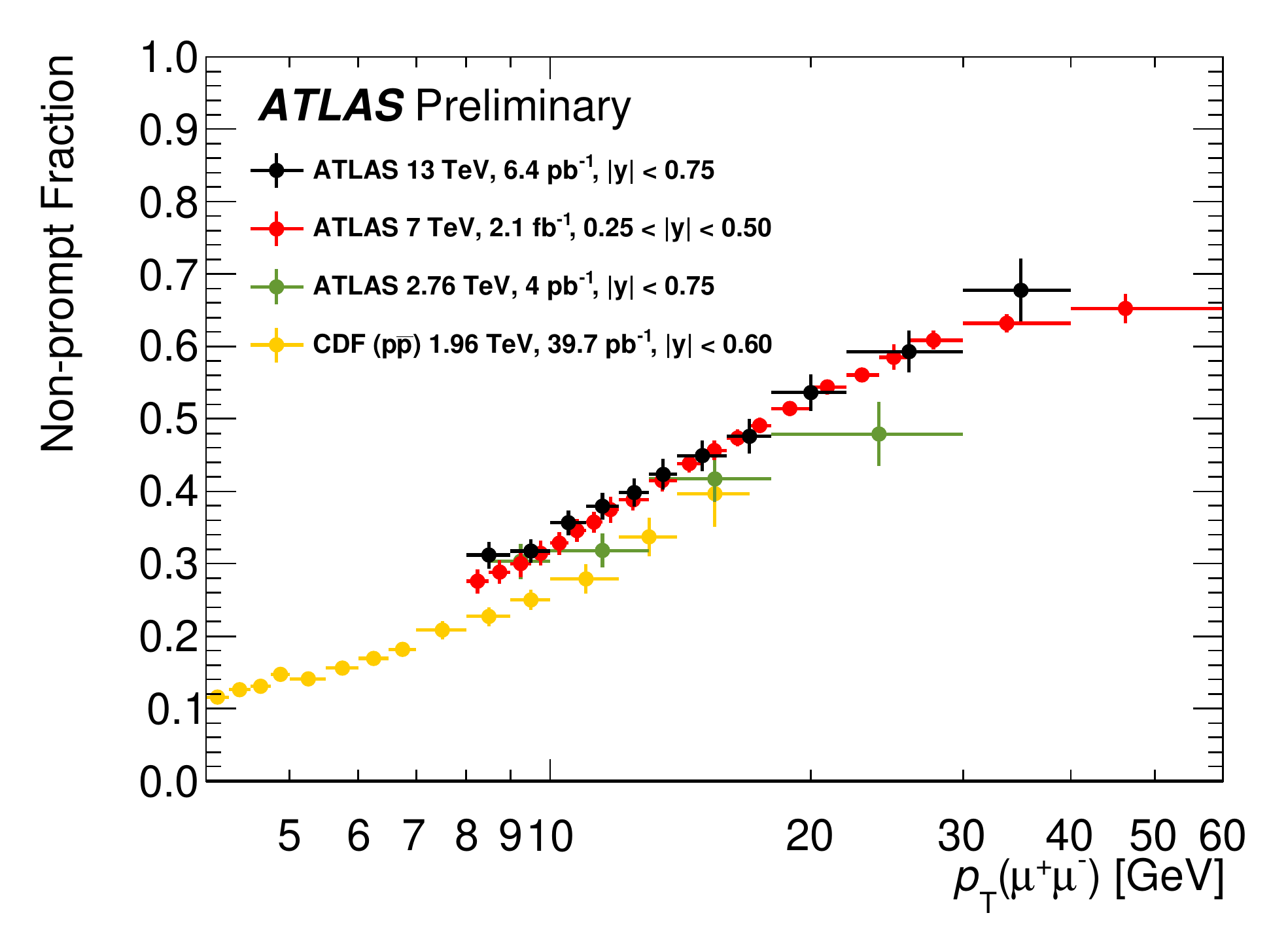}}
\caption{
The non-prompt fraction of $J/\psi$ production: (a) as a function of $p_T$ and rapidity at 7~TeV \cite{ATLAS:2015zdw} and (b)
its evolution with collision energy (see~\cite{TheATLAScollaboration:2015anj} and references therein).
}
\label{fig:frachistory}
\end{figure}

\begin{figure}[h]
%\begin{figure}[htbp!]
\centering
\subfloat[]{\includegraphics[height = 5.4cm, keepaspectratio]{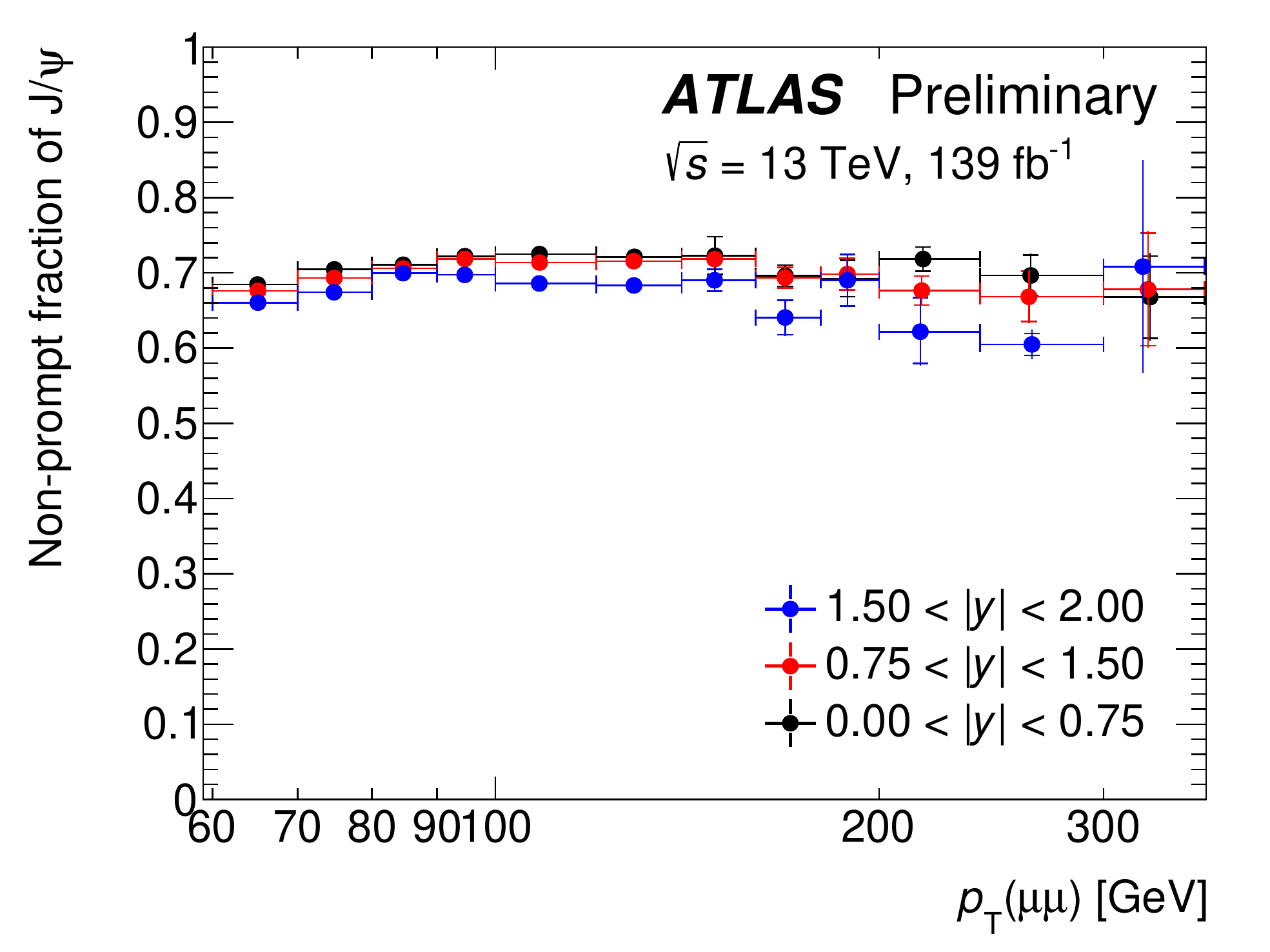}}
\subfloat[]{\includegraphics[height = 5.4cm, keepaspectratio]{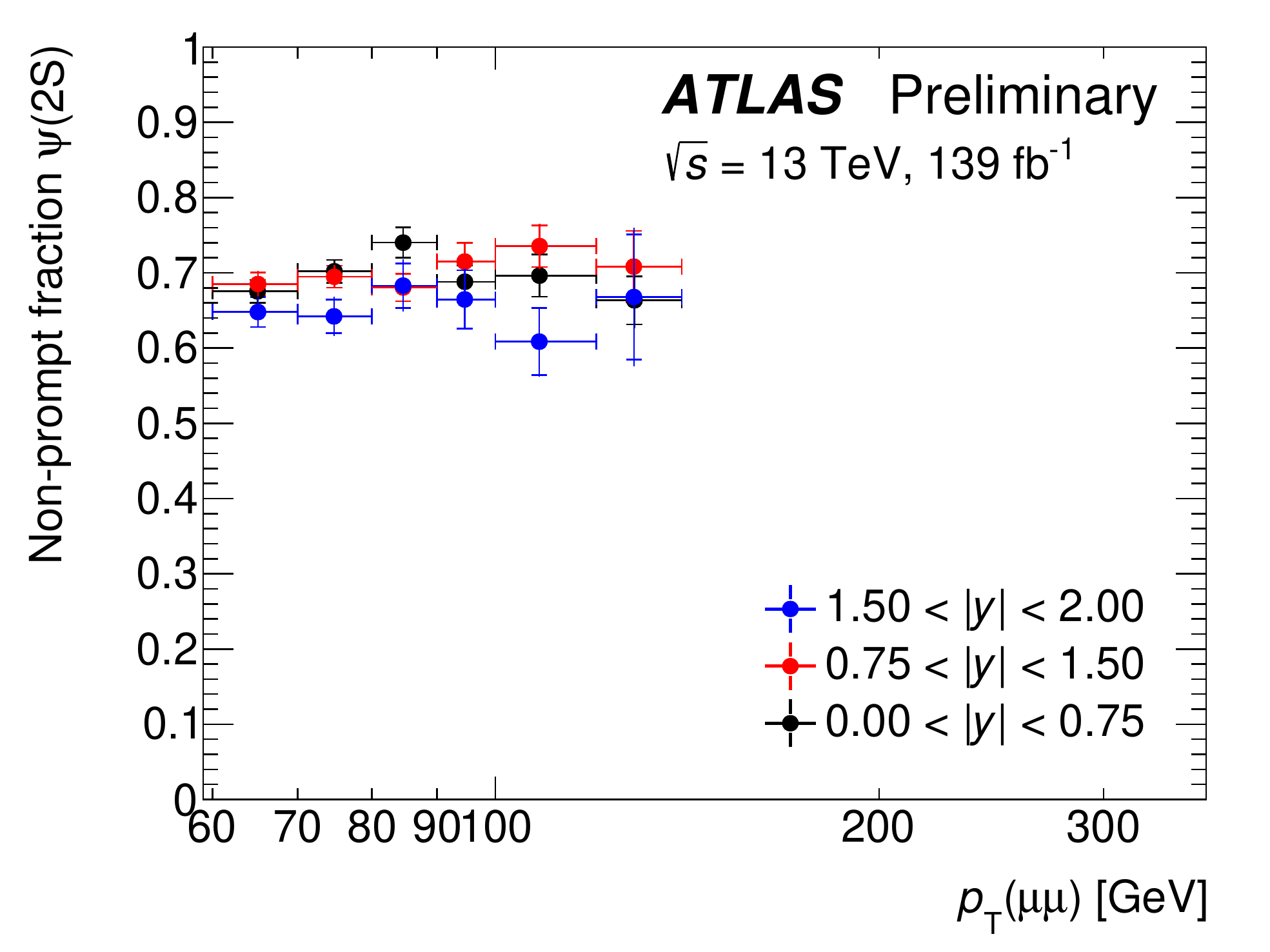}}
\caption{
Acceptance- and efficiency-corrected non-prompt production fractions for (a) $J/\psi$ and (b) $\psi(2S)$ mesons~\cite{ATLAS:2019ilf}. 
}
\label{fig:npfrac}
\end{figure}

Previous measurements of the non-prompt production fractions of $J/\psi$ and $\psi(2S)$ mesons have shown a weak dependence on rapidity but  a fairly strong dependence on $p_T$, with little evolution with collision energy from 2 to 13~TeV (Fig.~\ref{fig:frachistory}).
However, this measurement shows that in the high $p_T$ range above 60~GeV the non-prompt fraction stabilises and remains essentially constant at around 70\%, both for $J/\psi$ and $\psi(2S)$ (Fig.~\ref{fig:npfrac}). 

\begin{figure}[h]
\centering
\subfloat[]{\includegraphics[height = 5.4cm, keepaspectratio]{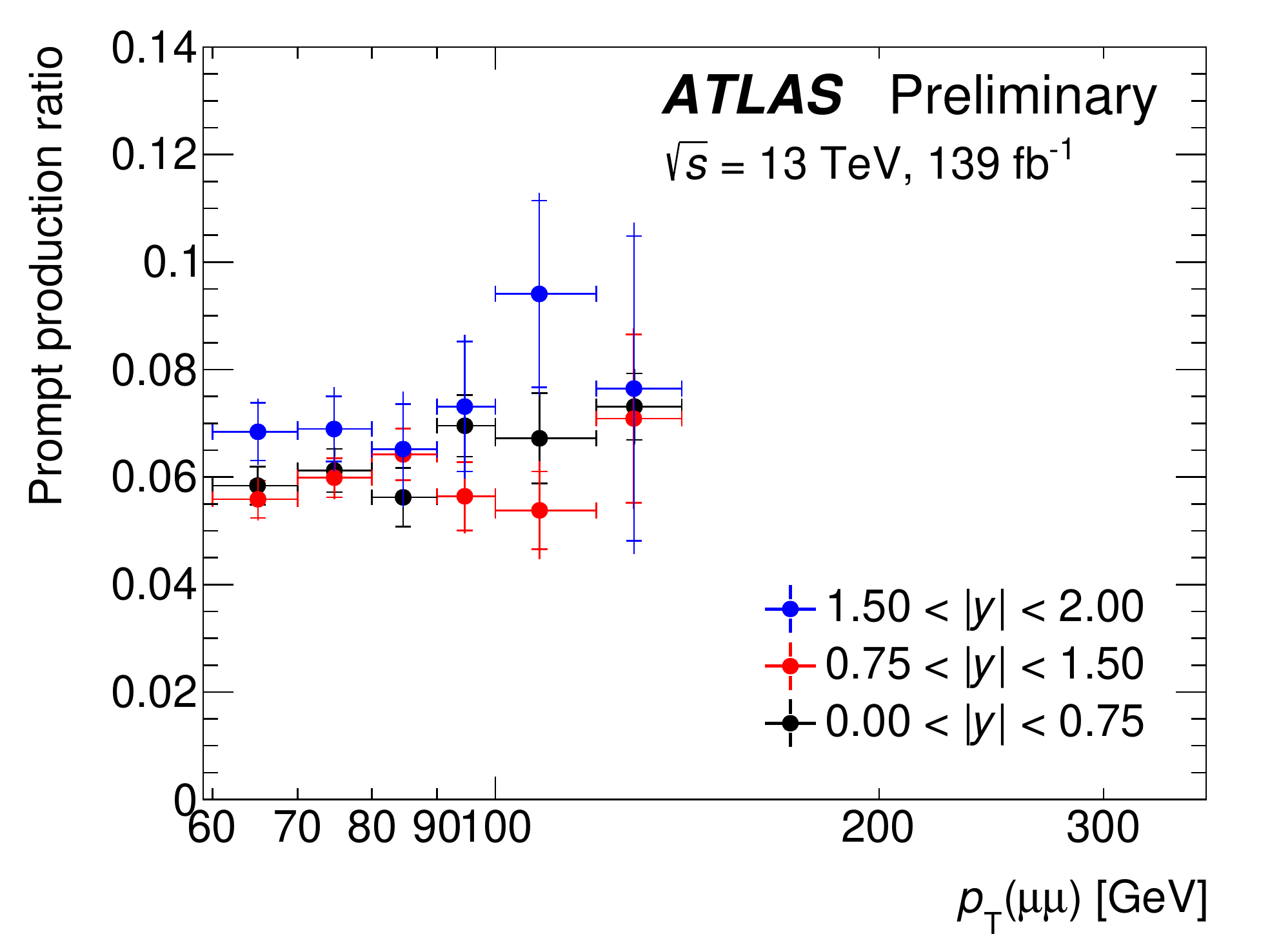}}
\subfloat[]{\includegraphics[height = 5.4cm, keepaspectratio]{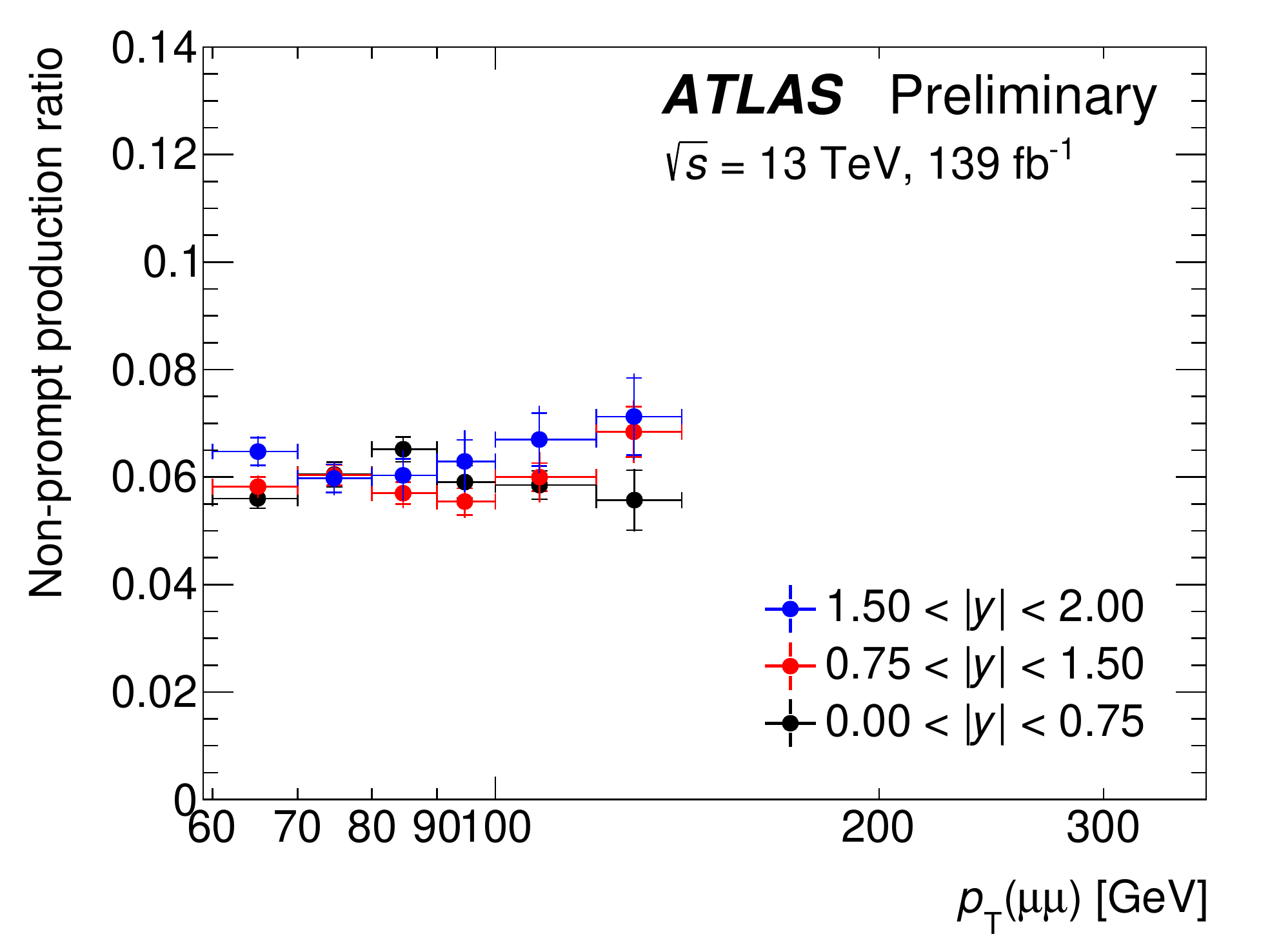}}
\caption{
Cross section ratios of $\psi(2S)$ over $J/\psi$, for (a) prompt and (b) non-prompt production mechanisms, after all relevant corrections were applied~\cite{ATLAS:2019ilf}. 
}
\label{fig:prodrat}
\end{figure}

\begin{figure}[h]
\centering
\subfloat[]{\includegraphics[height = 5.4cm, keepaspectratio]{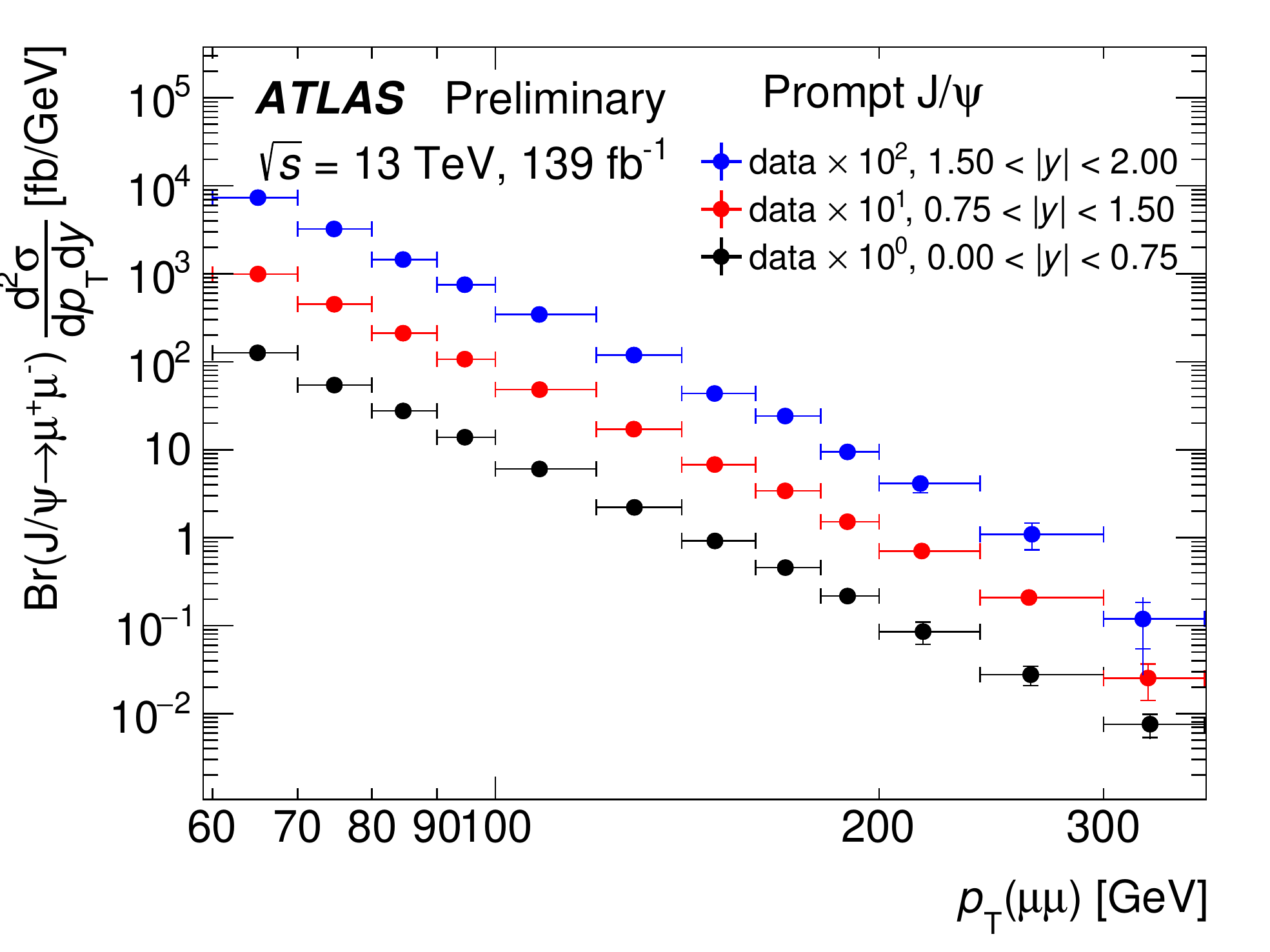}}
\subfloat[]{\includegraphics[height = 5.4cm, keepaspectratio]{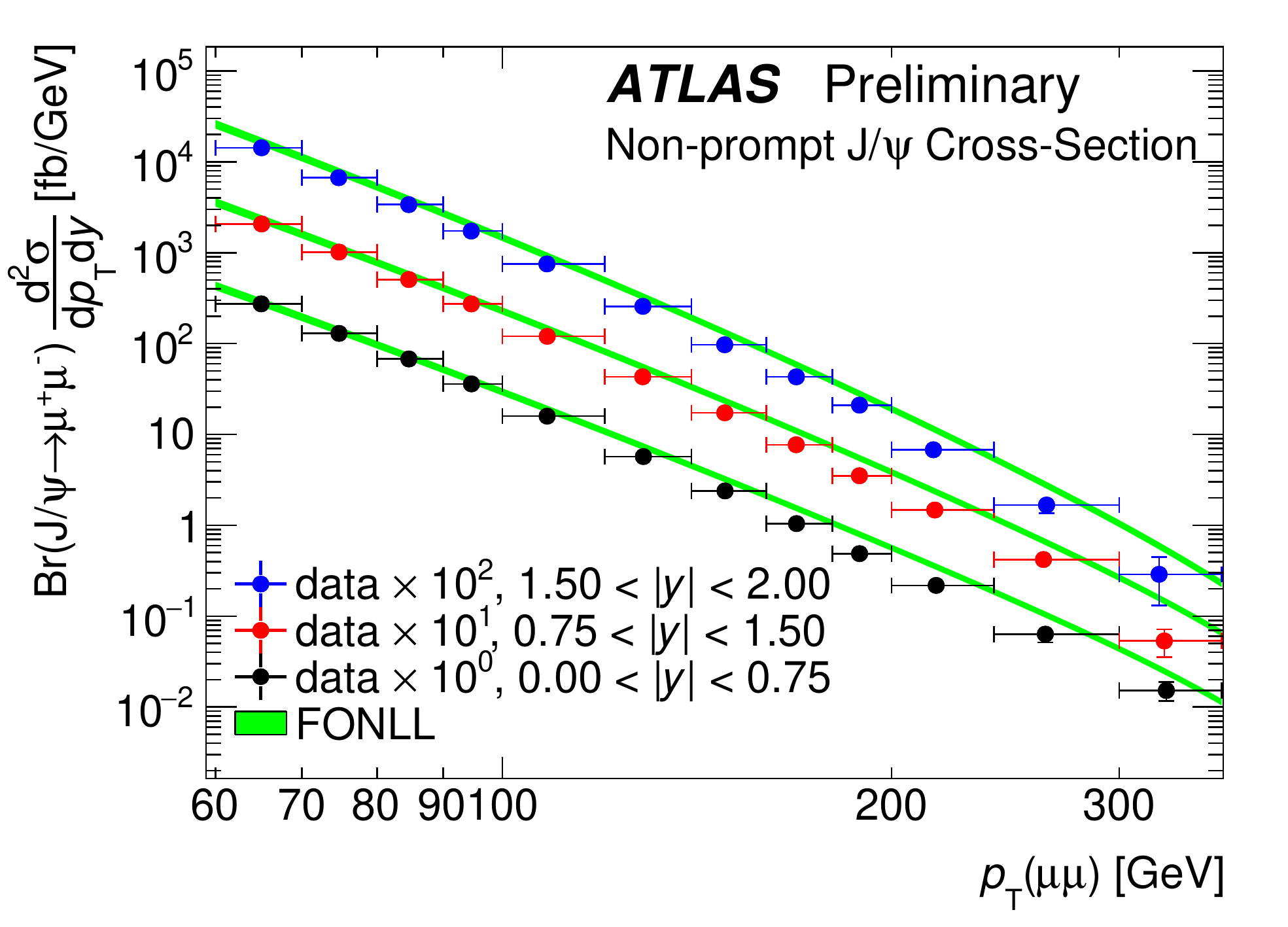}}
\caption{
Differential production cross sections of (a) prompt and (b) non-prompt $J/\psi$ mesons, in three slices of absolute rapidity ~\cite{ATLAS:2019ilf}. All relevant corrections have been applied. The green curves in (b) represent the predictions obtained by the FONLL model~\cite{Cacciari_2001,Cacciari:2012ny}. 
}
\label{fig:diffjpsi}
\end{figure}

The production ratios of $\psi(2S)$ over $J/\psi$ were measured separately for prompt and non-prompt production mechanisms, and showed similar results, relatively stable with $p_T$ and $y$ at the level of $\sim0.06$, as shown in Fig.~\ref{fig:prodrat}.
The differential cross sections, however, decrease rapidly with increasing $p_T$, continuing the trend from previous measurements. This was found to be the case for both prompt and non-prompt contributions, as shown in  Fig.~\ref{fig:diffjpsi} 
%and in Fig.~\ref{fig:diffpsi2s} for $\psi(2S)$.
for the $J/\psi$ case. 
\begin{comment}
\end{comment}
The non-prompt cross sections are described reasonably well by the FONLL model~\cite{Cacciari_2001,Cacciari:2012ny}, which deviates from the data by no more than a factor of 2 over five orders of magnitude of the cross section variation. Similar results were obtained for $\psi(2S)$, albeit in a narrower $p_T$ range of 60 to 140~GeV covered by the measurement (see ~\cite{ATLAS:2019ilf} for details).

\begin{figure}[h]
\centering
\includegraphics[width=0.6\textwidth]{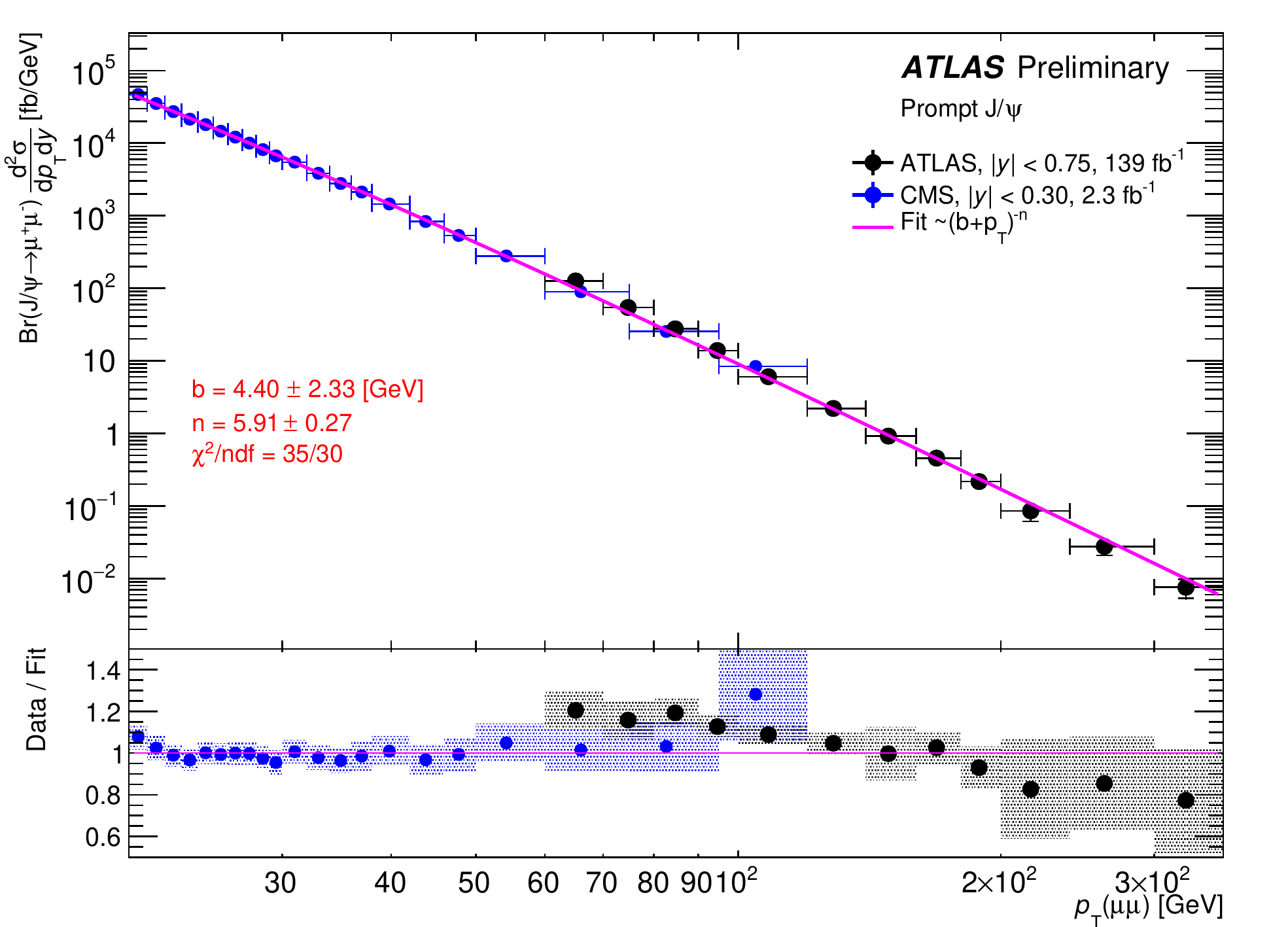}
\caption{
Comparison of the differential cross section of prompt $J/\psi$ production measured by ATLAS~\cite{ATLAS:2019ilf}  with the CMS results~\cite{CMS:2017dju}, in the central rapidity range.
}
\label{fig:cmsfitcomp}
\end{figure}

In the absence of specific predictions for prompt $J/\psi$ production in this $p_T$ range, a comparison is made with an earlier measurement from CMS~\cite{CMS:2017dju}, which covers an overlapping interval $20 < p_T < 150$~GeV. This is shown in Fig.~\ref{fig:cmsfitcomp}, 
together with a simple fit of the form $\sim(b+p_T)^{-n}$. This functional dependence seems to describe the data well in the whole range of $p_T$ between 20 and 360~GeV over seven orders of magnitude of the differential cross section variation, for $b=4.4$~GeV and $n\simeq6$. It would be interesting to 
see how NRQCD-based model calculations~\cite{PhysRevD.51.1125} compare with the data, especially at high $p_T$.

\section{Conclusion}
%You must include a conclusion.

In conclusion, heavy quarkonium studies remain an important part of the ATLAS physics programme, and this latest (preliminary) measurement provides new inputs for theoretical models, which should improve our understanding of QCD and related areas of particle physics. 

\section*{Acknowledgement}
These studies were supported in part
by  Shota Rustaveli National Science Foundation of Georgia (SRNSFG)  through the grant project \#DI-18-293.

\noindent
Copyright 2021 CERN for the benefit of the ATLAS Collaboration.                       
CC-BY-4.0 license.

% TODO:
% Provide your bibliography here. You have two options:

% FIRST OPTION - write your entries here directly, following the example below, including Author(s), Title, Journal Ref. with year in parentheses at the end, followed by the DOI number.

%\bibliographystyle{SciPost_bibstyle} % Include this style file here only if you are not using our template
%\bibliography{DIS_VK_charmonium}
%\bibliography{DIS_VK_charmonium.bib}
%\bibliography{SciPost_Example_BiBTeX_File.bib}

%\nolinenumbers

\end{document}